\documentclass{article}
\usepackage{spconf}
\usepackage{graphicx}
\usepackage{psfrag}
\usepackage{url}
\usepackage{stfloats}
\usepackage{amsmath,epsfig,amsfonts,amssymb,graphics,psfrag,theorem,calc,subfigure,url,bm,cite}
\usepackage{array}
\usepackage{caption}
\usepackage{calc}
\usepackage{subfig}
\usepackage{cases}
\usepackage{verbatim}
\usepackage{algorithm,algpseudocode}

\newtheorem{Lemma}{Lemma}

\newtheorem{Theorem}{Theorem}

\theorembodyfont{\rmfamily}

{\begin{list}{}{
    \settowidth{\labelwidth}{\mbox{\textnormal{#1}}}%
    \setlength{\leftmargin}{\labelwidth+\labelsep}}}%
{\end{list}}

\title{A Robust Artificial Noise Aided transmit design for MISO Secrecy}
\name{Qiang Li and Wing-Kin Ma\thanks{This work is partially
supported by a General Research Fund of Hong Kong Research Grant
Council (CUHK415908).}}

\address {Department of
Electronic Engineering,
The Chinese University of Hong Kong, \\ Shatin, N.T., Hong Kong \\
\small E-mail:qli@ee.cuhk.edu.hk, wkma@ieee.org}

\begin{document}
\ninept
%

\maketitle
\begin{abstract}
This paper considers an artificial noise (AN) aided secrecy rate
maximization (SRM) problem for a multi-input single-output (MISO)
channel overheard by multiple single-antenna
eavesdroppers. We assume that the transmitter has perfect knowledge
about the channel to the desired user but imperfect knowledge about
the channels to the eavesdroppers. Therefore, the resultant SRM
problem is formulated in the way that we maximize the worst-case
secrecy rate by jointly designing the signal covariance ${\bf W}$
and the AN covariance ${\bf \Sigma}$. However, such a worst-case SRM
problem turns out to be hard to optimize, since it is nonconvex in
${\bf W}$ and ${\bf \Sigma}$ jointly. Moreover, it falls into the
class of semi-infinite optimization problems. Through a careful
reformulation, we show that the worst-case SRM problem can be
handled by performing a one-dimensional line search in which a
sequence of semidefinite programs (SDPs) are involved. Moreover, we
also show that the optimal ${\bf W}$ admits a rank-one structure,
implying that transmit beamforming is secrecy rate optimal under the
considered scenario. Simulation results are provided to
demonstrate the robustness and effectiveness of the proposed design
compared to a non-robust AN design.
\end{abstract} \vspace{-0.0cm}


\section{Introduction}
In the last decade, multi-antenna techniques have been extensively
investigated from the perspective of providing high throughput.
Recently, there has been much interest in using multiple antennas to
achieve secure communication, which is known as {\it physical-layer
secrecy}. In a traditional single-input single-output scenario, the
idea of physical-layer secrecy is to add some structured redundancy
in the transmitted signal such that the desired user can correctly
decode the confidential information, but for the eavesdropper he/she
cannot retrieve anything from the observation~\cite{Liang_book}. To
make physical-layer secrecy viable, a prerequisite is that the
desired user's channel has to be better than the eavesdropper's.
However, this may not be satisfied if the transmitter
has only a single antenna; e.g., when the eavesdropper is closer to
the transmitter with lower reception noise power than the desired
user. To alleviate the dependence of the channels, recent studies
are mainly focused on employing multiple antennas to transmit, since
multiple transmit antennas provide additional spatial degree of
freedom to degrade the reception of the eavesdropper. A possible way
to do this is transmit beamforming, which premultiplies the signal
by a weight vector such that the power radiation is concentrated
over the direction of the desired user. In addition to concentrating
the signal power, a more active way is to use part of the power to
artificially generate some noise to interfere the eavesdropper. This
artificial noise (AN) approach was first proposed
in~\cite{Negi2005}, and has been shown to be effective in improving secrecy rates~\cite{KW2007,Negi2005,Swindlehurst2009,Mukherjee2009,XYZHOU,Jorswieck}.

Current studies on AN aided transmit design are mainly based on a
premise that AN lies in the orthogonal complement subspace of the
desired user's channel in an isotropic fashion, e.g.,~\cite{KW2007,Negi2005,Swindlehurst2009,Mukherjee2009,XYZHOU}.
This isotropic AN design has an advantage that no eavesdropper's
channel state information (CSI) is needed at the transmitter,
thereby making it very suitable for the passive eavesdropper scenario. On
the other hand, as demonstrated in~\cite{Swindlehurst2009,Liao10},
when the eavesdropper's CSI is perfectly known at the transmitter,
e.g., when the eavesdropper is also a user of the system, we can
block the eavesdropper more effectively by aligning AN to the eavesdropper's direction through judicious optimization, rather than keeping the AN
isotropically. However this perfect CSI assumption may be too
stringent. Thus, in this paper we consider a scenario where the
transmitter has only imperfect CSIs of the eavesdroppers, and we attempt to
maximize the worst-case secrecy rate by jointly optimizing the
signal and the AN covariances. Through a careful
reformulation, we show that the AN aided worst-case secrecy rate
maximization (SRM) problem can be handled by performing a
one-dimensional line search in which a sequence of semidefinite
programs (SDPs) are involved. Moreover, we prove that the
optimal transmit covariance for the signal part admits a rank-one
structure, and thus, transmit beamforming is an optimal transmit
strategy for the scenario considered.

There are some related works worth
mentioning. In~\cite{Swindlehurst2009,Liao10}, AN
aided transmit beamforming designs are considered from a QoS
perspective. Specifically, they focus on the
signal-to-interference-and-noise-ratio (SINR) at each receiver
(including the eavesdroppers), instead of the secrecy rate
considered here. Moreover, this work considers imperfect CSI in a
worst-case sense. In~\cite{Jorswieck}, the authors consider the AN
design for secrecy rate maximization under the stochastic CSI
uncertainty model, as opposed to the worst-case deterministic model considered
here.

{\it Notation}: $\mathbf{A}^H$, $\text{Tr}(\mathbf{A})$ and
rank(${\bf A}$) represent Hermitian transpose, trace and rank of a
matrix $\mathbf{A}$; $\mathbf{I}$ is an identity matrix of
appropriate size; $\mathbf{A}\succeq \mathbf{0}$ $({\bf A}\succ {\bf
0})$ means $\mathbf{A}$ is Hermitian positive semidefinite
(definite) matrix; $\mathbb{H}^{N}$ denotes the set of $N$-by-$N$
Hermitian matrices; ${\bf x} \sim \mathcal{CN} ({\bm \mu}, {\bm
\Sigma})$ means that ${\bf x}$ is a random vector following a
complex circular Gaussian distribution with mean ${\bm \mu}$ and
covariance $\bf \Sigma$; $\text{E}\{\cdot\}$ is the expectation
operator.

\section{System Model and Problem Statement}
We consider a wireless network that consists of a
single transmitter with $N_t$ antenna elements, and $K+1$ single
antenna receivers. Among these $K+1$ receivers, only one receiver is
legitimate while the others are eavesdroppers. The transmitter
attempts to send a confidential message to the legitimate receiver.
For convenience, we refer to the transmitter, legitimate receiver, and
the eavesdroppers as {\it Alice}, {\it Bob}, and {\it Eves},
respectively. Let ${\bf h}$ and ${\bf g}_k$ denote $N_t \times 1$
complex channel vectors from Alice to Bob and the $k$th Eve,
respectively. Then, the received signal can be expressed as
\begin{subequations} \label{eq:sig_mod}
\begin{align}
y_{b}(t) & = {\bf h}^H {\bf x}(t) + n(t), \\
y_{e,k}(t) & = {\bf g}_k^H {\bf x}(t) + v_k(t),\quad k=1,\ldots,K.
\end{align}
\end{subequations}
where $n(t) \in \mathbb{C}$ and $v_k(t)\in \mathbb{C}$ are
independent additive white Gaussian noises with mean zero and unit
variance; ${\bf x}(t) \in \mathbb{C}^{N_t}$ is the transmit signal
vector, which possesses the following form
\[ {\bf x}(t) = {\bf s}(t) + {\bf z}(t). \]
Here, ${\bf s}(t)$ is the confidential information intended for Bob and we define its covariance as ${\bf W} = {\rm E} \{{\bf s}(t) {\bf s}^H(t)\} $;
${\bf z}(t)$ is the noise vector artificially created by Alice to
interfere Eves. In this work, we assume that ${\bf z}(t)$ and ${\bf s}(t)$ are independent, and ${\bf z}(t) \sim
\mathcal{CN} ({\bf 0}, {\bm \Sigma})$. Generally speaking, we wish to design ${\bf W}$ and ${\bf \Sigma}$ so that a good information secrecy can be achieved.

To describe the imperfect CSI model for Eves, we
assume that Alice knows only channel estimates of ${\bf
g}_k$, i.e.,
\begin{equation}
  {\bf g}_k = \bar{\bf g}_k + \Delta {\bf g}_k,~k=1,\ldots,K,
\end{equation}
where $\bar{\bf g}_k$ is the channel estimate at Alice; $\Delta {\bf
g}_k$ represents the channel uncertainty. These uncertainties are
assumed to be deterministic unknowns with bounds on their
magnitudes:
\[ \| \Delta {\bf g}_k \|_2 \leq \epsilon_k,~~k=1,\ldots,K \]
for some $\epsilon_1,\ldots,\epsilon_K > 0$.

Under the above described uncertainty model, the proposed robust SRM
formulation is given by~\cite{Liang2007}
\vspace{10pt}

\hspace{-1cm}
\begin{minipage}[h]{0.49\textwidth}
\fbox{
\parbox{\textwidth}{
\begin{equation}\label{eq:rob_SRM}
\begin{aligned}
R^\star(P) =  \max_{ {\bf W}, {\bm \Sigma} } ~~&
 \Big \{ \min_{k=1,\ldots,K} f_{k} ({\bf W}, {\bm \Sigma})  \Big \} \\
{\rm s.t.} ~~& {\rm Tr}({\bf W} + {\bf \Sigma} ) \leq P,\\
~~ &  {\bf W} \succeq {\bf 0}, \quad {\bm \Sigma} \succeq {\bf 0},
\end{aligned}
\end{equation}
where \[
f_{k} ({\bf W}, {\bm \Sigma})  =  \log \Big(1+
\frac{ {\bf h}^H {\bf W} {\bf h} }{1 + {\bf h}^H {\bm \Sigma} {\bf
h}} \Big) - \max_{{\bf g}_k \in {\mathcal B}_k } \log \Big( 1 +
\frac{ {\bf g}_k^H {\bf W} {\bf g}_k } {1 + {\bf g}_k^H {\bm \Sigma}
{\bf g}_k } \Big), \]
\[ {\mathcal B}_k  = \{ {\bf g}_k
\in {\mathbb C}^{N_t} ~|~ \| {\bf g}_k - {\bar {\bf g}}_k \|_2 \leq
\epsilon_k \}. \]
}
}
\end{minipage}

\vspace{10pt}

The function $f_k(\cdot)$ represents the worst secrecy rate function
among all channel possibilities, for the $k$th Eve. Moreover, in arriving at~\eqref{eq:rob_SRM}, ${\bf s}(t)$
is assumed to be Gaussian distributed. It should be
noted that the original SRM formulation in~\cite{Liang2007} assumes
perfect CSIs at Alice. Herein~\eqref{eq:rob_SRM} can be regarded as
a robust counterpart of~\cite{Liang2007}, which guarantees that the
worst secrecy rate is no less than $R^\star(P)$ for any channel
possibilities (described by ${\mathcal B}_k$).

The aim of this paper is to propose a method to jointly
optimize ${\bf W}$ and ${\bf \Sigma}$ for~\eqref{eq:rob_SRM}.
However, it is challenging to do so. This is because
Problem~\eqref{eq:rob_SRM} is nonconvex in ${\bf W}$ and ${\bf
\Sigma}$ jointly, even under the situation $K=1$ and $\epsilon_1 =
0$. In the following section, we will develop a tractable
approach to tackling Problem~\eqref{eq:rob_SRM}.

\section{A Tractable Approach to the Robust SRM }
To describe the proposed approach, let us rewrite \eqref{eq:rob_SRM}
as
\begin{subequations}\label{eq:rob_SRM_eqv1}
\begin{align}
 \max_{ {\bf W}, {\bm \Sigma}, \beta } & ~
\log \Big(1+ \frac{ {\bf h}^H {\bf W} {\bf h} }{1 + {\bf h}^H {\bm
\Sigma} {\bf
h}} \Big) - \log \beta \label{eq:rob_SRM_eqv1_a}\\
{\rm s.t.} & ~ \max_{{\bf g}_k \in {\mathcal B}_k} \log \Big(1+
\frac{ {\bf g}_k^H {\bf W} {\bf g}_k }{1 + {\bf g}_k^H {\bm \Sigma}
{\bf
g}_k } \Big) \leq \log \beta,~\forall k \label{eq:rob_SRM_eqv1_b}\\
& ~ {\rm Tr}({\bf W} + {\bf \Sigma} ) \leq P, \quad {\bf W} \succeq
{\bf 0}, \quad {\bm \Sigma} \succeq {\bf
0},\label{eq:rob_SRM_eqv1_c}
\end{align}
\end{subequations}
where the slack variable $\beta$ is introduced to simplify the
objective function. By the monotonicity of $\log$ function, we
simplify \eqref{eq:rob_SRM_eqv1} as
\begin{subequations}\label{eq:rob_SRM_eqv2}
\begin{align}
 \max_{ {\bf W}, {\bm \Sigma}, \beta } & ~ \frac{ 1 +
{\bf h}^H ( {\bf W} + {\bf \Sigma} ) {\bf h} }{\beta ( 1 + {\bf h}^H
{\bf \Sigma} {\bf
h} )} \label{eq:rob_SRM_eqv2_a}  \\
{\rm s.t.} & ~ \max_{{\bf g}_k \in {\mathcal B}_k} {\bf
g}_k^H \big( {\bf W} - (\beta -1){\bf \Sigma} \big ) {\bf g}_k \leq \beta -1,~\forall k \label{eq:rob_SRM_eqv2_b}\\
& ~ {\rm Tr}({\bf W} + {\bf \Sigma}) \leq P, \quad {\bf W} \succeq
{\bf 0}, \quad {\bm \Sigma} \succeq {\bf
0}.\label{eq:rob_SRM_eqv2_c}
\end{align}
\end{subequations}
Problem~\eqref{eq:rob_SRM_eqv2} is a nonconvex and semi-infinite
problem due to \eqref{eq:rob_SRM_eqv2_a} and
\eqref{eq:rob_SRM_eqv2_b}. Let us simplify \eqref{eq:rob_SRM_eqv2_a}
first. The idea is to apply the Charnes-Cooper
transformation~\cite{Charnes1962}. To do so, we denote
\begin{equation}\label{eq:charnes_copper}
{\bf W} = {{\bf Z}/ \xi},~~{\bf \Sigma} = {\bf Q}/ \xi
\end{equation}
for some ${\bf Z} \succeq {\bf 0}, {\bf Q} \succeq {\bf 0},$ and
$\xi
>0$. Problem~\eqref{eq:rob_SRM_eqv2} can be equivalently written as
\begin{subequations}\label{eq:rob_SRM_eqv3}
\begin{align}
 \max_{ {\bf Z}, {\bf Q}, \beta, \xi } & ~
 \frac{ \xi + {\bf h}^H ( {\bf Z} + {\bf Q} ) {\bf h}
}{\beta ( \xi + {\bf h}^H {\bf Q} {\bf
h} )} \label{eq:rob_SRM_eqv3_a}  \\
{\rm s.t.} & ~ \max_{{\bf g}_k \in {\mathcal B}_k} {\bf
g}_k^H \big( {\bf Z} - (\beta -1){\bf Q} \big ) {\bf g}_k \leq (\beta -1) \xi,~\forall k \label{eq:rob_SRM_eqv3_b}\\
& ~{\rm Tr}({\bf Z} + {\bf Q}) \leq \xi P, \quad {\bf Z} \succeq
{\bf 0}, \quad {\bf Q} \succeq {\bf 0}, \quad \xi > 0,
\label{eq:rob_SRM_eqv3_c}
\end{align}
\end{subequations}
which can be further reformulated as
\begin{subequations}\label{eq:rob_SRM_eqv4}
\begin{align}
 \max_{ {\bf Z}, {\bf Q}, \beta, \xi } & ~
\xi + {\bf h}^H ( {\bf Z} + {\bf Q} ) {\bf h} \label{eq:rob_SRM_eqv4_a}  \\
{\rm s.t.} & ~{\beta ( \xi + {\bf h}^H {\bf Q} {\bf h} )} = 1 \label{eq:rob_SRM_eqv4_b}\\
& ~ \max_{{\bf g}_k \in {\mathcal B}_k} {\bf
g}_k^H \big( {\bf Z} - (\beta -1){\bf Q} \big ) {\bf g}_k \leq (\beta -1) \xi,~\forall k \label{eq:rob_SRM_eqv4_c}\\
& ~ {\rm Tr}({\bf Z} + {\bf Q}) \leq \xi P, \quad {\bf Z} \succeq
{\bf 0}, \quad {\bf Q} \succeq {\bf 0}, \quad \xi \geq 0.
\label{eq:rob_SRM_eqv4_d}
\end{align}
\end{subequations}
Here \eqref{eq:rob_SRM_eqv4_b} is introduced to fix the denominator
of \eqref{eq:rob_SRM_eqv3_a}; $\xi > 0$ in \eqref{eq:rob_SRM_eqv3_c}
is replaced with $\xi \geq 0$ in \eqref{eq:rob_SRM_eqv4_d}. This
replacement does not cause any problem to the equivalence since any
feasible $\xi$ of \eqref{eq:rob_SRM_eqv4} must be positive;
otherwise \eqref{eq:rob_SRM_eqv4_b} and \eqref{eq:rob_SRM_eqv4_d}
cannot be met simultaneously. Readers are referred to~\cite{QLI2010} for the detailed argument.

Problem~\eqref{eq:rob_SRM_eqv4} is still hard to optimize due to the
semi-infinite constraint \eqref{eq:rob_SRM_eqv4_c}. In particular,
we have to check that \eqref{eq:rob_SRM_eqv4_c} is satisfied for
every possible ${\bf g}_k$ in $\mathcal{B}_k$, which is
computationally prohibitive. To make \eqref{eq:rob_SRM_eqv4_c} more
tractable, we need the following $\mathcal{S}$-procedure, which will
be used to transform \eqref{eq:rob_SRM_eqv4_c} into a matrix inequality:
\begin{Lemma}[\cite{Boyd2004}]
Let
\[ \varphi_k({\bf x}) = {\bf x}^H {\bf A}_k {\bf x} + 2 {\rm Re}\{ {\bf b}_k^H {\bf x} \} + c_k \]
for $k=1,2$, where ${\bf A}_k \in \mathbb{H}^n$, ${\bf b}_k \in
\mathbb{C}^n$, $c_k \in \mathbb{R}$. The implication $\varphi_1({\bf x})
\leq 0 \Rightarrow \varphi_2({\bf x}) \leq 0$ holds if and only if there
exists $\mu \geq 0$ such that
\[  \mu \begin{bmatrix} {\bf A}_1 & {\bf b}_1 \\ {\bf b}_1^H & c_1 \end{bmatrix}
- \begin{bmatrix} {\bf A}_2 & {\bf b}_2 \\ {\bf b}_2^H & c_2
\end{bmatrix} \succeq {\bf 0}, \] provided that there exists a point
$\hat{\bf x}$ such that $\varphi_1(\hat{\bf x}) < 0$.
\end{Lemma}
Now consider~\eqref{eq:rob_SRM_eqv4_c}. It can interpreted as the
following implication:\vspace{-5pt}
\begin{equation}
\begin{aligned}
  \Delta {\bf g}_k^H \Delta{\bf g}_k - \epsilon_k^2 \leq 0
\Rightarrow & \Delta {\bf g}_k^H {\bf M } \Delta{\bf g}_k + 2 {\rm
Re} \{ \bar{\bf g}_k^H {\bf M} \Delta {\bf g}_k \} \\ & + \bar{\bf
g}_k^H {\bf M} \bar{\bf g}_k - (\beta - 1) \xi \leq 0,~~\forall k
\end{aligned}\vspace{-5pt}
\end{equation}
where ${\bf M} = {\bf Z} - (\beta -1) {\bf Q}$. By Lemma~1, we can
rewrite the above implication equivalently as the following matrix inequality:
\begin{equation}\label{eq:LMI}
\begin{aligned}
  & {\bf T}_k({\bf Z}, {\bf Q}, \beta, \mu_k,\xi) = \\
  & \begin{bmatrix}
    \mu_k {\bf I} - {\bf M} & -{\bf M} \bar{\bf g}_k \\
    -\bar{\bf g}_k^H {\bf M} & -\epsilon_k^2 \mu_k - \bar{\bf g}_k^H
    {\bf M} \bar{\bf g}_k + (\beta -1)\xi
  \end{bmatrix}  \succeq {\bf 0},
  \end{aligned}
\end{equation}
with $\mu_k \geq 0$. Substituting \eqref{eq:LMI} into \eqref{eq:rob_SRM_eqv4}, we
have\vspace{-5pt}
\begin{subequations}\label{eq:rob_SRM_eqv}
\begin{align}
 \max_{ {\bf Z}, {\bf Q}, \beta, \xi, {\bm \mu} } & ~
\xi + {\bf h}^H ( {\bf Z} + {\bf Q} ) {\bf h} \label{eq:rob_SRM_eqv_a}  \\
{\rm s.t.} & ~{\beta ( \xi + {\bf h}^H {\bf Q} {\bf h} )} = 1 \label{eq:rob_SRM_eqv_b}\\
& ~ {\bf T}_k({\bf Z}, {\bf Q}, \beta, \mu_k, \xi) \succeq {\bf 0},\quad \mu_k \geq 0,~~ \forall k\label{eq:rob_SRM_eqv_c}\\
& ~ {\rm Tr}({\bf Z} + {\bf Q}) \leq \xi P, ~~ {\bf Z} \succeq {\bf
0}, ~~ {\bf Q} \succeq {\bf 0}, ~~ \xi \geq 0.
\label{eq:rob_SRM_eqv_d}
\end{align}
\end{subequations}
While~\eqref{eq:rob_SRM_eqv} has a much simpler form compared
to~\eqref{eq:rob_SRM}, the former is still a nonconvex problem,
arising from the nonconvex constraints \eqref{eq:rob_SRM_eqv_b} and
\eqref{eq:rob_SRM_eqv_c}. However, if we fix $\beta$ in
\eqref{eq:rob_SRM_eqv}, then~\eqref{eq:rob_SRM_eqv} is just
an SDP problem. Motivated by this, we recast~\eqref{eq:rob_SRM_eqv}
in the following form:
\vspace{3pt}

\hspace{-0.55cm}
\begin{minipage}[h]{0.46\textwidth}
\fbox{
\parbox{\textwidth}{
\begin{subequations}\label{eq:s1}
  \begin{align}
\max_{\beta} & ~~ \phi(\beta) \label{eq:s1_a}\\
 {\rm s.t.} & ~~1 \leq \beta \leq 1+P\|{\bf h}\|^2, \label{eq:s1_b}
  \end{align}
\end{subequations}
where
\begin{subequations}\label{eq:s2}
  \begin{align}
\hspace{-5pt}\phi(\beta) = \max_{ {\bf Z}\succeq {\bf 0}, {\bf Q}\succeq {\bf 0},
\xi \geq 0, {\bm \mu}\geq {\bf 0 } }& ~~
\xi + {\bf h}^H ( {\bf Z} + {\bf Q} ) {\bf h} \label{eq:s2_a}  \\
 {\rm s.t.}&  ~~{\beta ( \xi + {\bf h}^H {\bf Q} {\bf h} )} = 1 \label{eq:s2_b}\\
&  ~~ {\bf T}_k({\bf Z}, {\bf Q}, \beta, \mu_k, \xi) \succeq {\bf 0},~\forall k \label{eq:s2_c}\\
&  ~~ {\rm Tr}({\bf Z} + {\bf Q}) \leq \xi P.
  \end{align}
\end{subequations}}
}
\end{minipage}

\vspace{3pt}
In~\eqref{eq:s1_b}, $\beta \geq 1$ follows from
\eqref{eq:rob_SRM_eqv1_b}; $\beta \leq 1+P\|{\bf h}\|^2$ is derived
from the following relation
\[  \beta \leq 1+\frac{{\bf h}^H {\bf W} {\bf h}}{1 + {\bf h}^H {\bf \Sigma} {\bf h}} \leq 1+{\bf h}^H {\bf W} {\bf h} \leq 1 + P \|{\bf h}\|^2,\]
where the leftmost inequality is a consequence of
\eqref{eq:rob_SRM_eqv1_a} and the requirement of nonnegative secrecy
rate (or the objective value of \eqref{eq:rob_SRM_eqv1} is
nonnegative); the rightmost inequality follows from
${\rm Tr}({\bf W}) \leq P$ and the equality holds when ${\bf W} = P {\bf h} {\bf
h}^H / \|{\bf h}\|^2$.

Since Problem~\eqref{eq:s1} is a single variable optimization
problem, we can perform a one-dimensional line search over $\beta$.
Such a search process involves computation of $\phi(\beta)$, which
is obtained by solving~\eqref{eq:s2}. As \eqref{eq:s2} is an SDP, it
can be efficiently and reliably solved by available softwares, e.g.,
\verb"CVX"~\cite{Grant2009}. After completing the search, we can
easily obtain the transmit covariances ${\bf W}$ and ${\bf \Sigma}$
for our main problem~\eqref{eq:rob_SRM} through the
relation~\eqref{eq:charnes_copper}.

Before we close this section, here is one remaining issue worth
investigating---what structure the optimal ${\bf W}^\star$
should possess, or more precisely, what the rank property of ${\bf
W}^\star$ should have. The following theorem gives the
answer:\vspace{-5pt}
\begin{Theorem}
  Suppose that $R^\star(P)> 0$\footnote{If $R^\star(P) = 0$, then, apparently, ${\bf W}^\star = {\bf \Sigma}^\star = {\bf 0}$ is optimal to \eqref{eq:rob_SRM}.}. Then, there exists
  an optimal ${\bf W}^\star$ of
  \eqref{eq:rob_SRM} such that ${\rm rank}({\bf W}^\star) = 1$.
\end{Theorem}\vspace{-5pt}
The proof of Theorem~1 is given in the Appendix. A key ingredient of
proving Theorem~1 is to consider a secrecy rate related power
minimization problem and investigate its Karush-Kuhn-Tucker (KKT)
conditions. Also note that the proof is constructive and hence the rank-one ${\bf W}^\star$ can be found in practice. Theorem~1 provides a useful physical-layer design
guideline that {\it transmit beamforming} is an optimal transmit
strategy for the artificial noise aided secure transmission under
the scenario considered. \vspace{-5pt}

\section{Simulation Results and Conclusions}
\begin{figure}[!h]
\centerline{\resizebox{.42\textwidth}{!}{\includegraphics{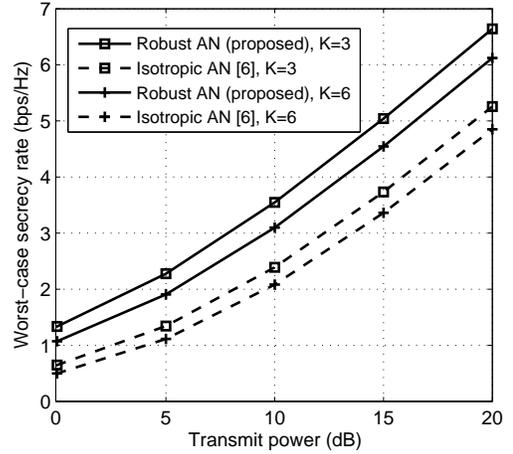}}
} \vspace*{-.5\baselineskip} \caption{Worst-case secrecy rate versus
transmit power. $N_t=4$, $\alpha=0.1$. } \label{fig:example1}
\vspace*{-1.1\baselineskip}
\end{figure}

\begin{figure}[!h]
\centerline{\resizebox{.42\textwidth}{!}{\includegraphics{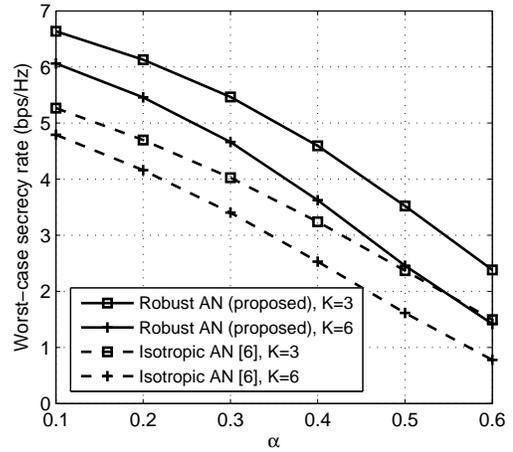}}
} \vspace*{-.5\baselineskip} \caption{Worst-case secrecy rate versus
channel uncertainty. $N_t=4$, $P=20$dB.} \label{fig:example2}
\vspace*{-1.5\baselineskip}
\end{figure}

\vspace{-8pt}
We provide two simulation examples to test the performance of the
robust AN design proposed in Section~3 and compare it with an equal-power-splitting isotropic AN design~\cite{XYZHOU}, which splits half of the power to transmit the
confidential information over the direction of ${\bf h}$, while the
remaining half, as artificial noise, is spread isotropically in the
orthogonal complement subspace of ${\bf h}$. In the following
simulations, we denote the normalized channel uncertainty $\alpha_k
= \epsilon_k/\sqrt{{\rm E}\{ \| {\bf g}_k\|^2 \}},~\forall k$ and set
$\alpha_1=\ldots,\alpha_K = \alpha$, i.e., the same uncertainty
level for all Eves' channel links. The elements of $\bf{h}$ and
${\bf g}_k$ are i.i.d. complex Gaussian distributed with mean 0 and
variance 1. All results were averaged over 1000 independent channel
realizations. Fig.~\ref{fig:example1} evaluates the relationship
between the worst-case secrecy rate (i.e., the objective value in~\eqref{eq:rob_SRM}) and the transmit power level for
different number of Eves. It can be seen from the figure that the
proposed robust AN design outperforms the non-robust AN design over
the whole power range tested. In particular, for $P=20$dB, $K=3$ and
$\alpha = 0.1$, the worst-case secrecy rate gap between these two
designs is about $1.5$ bps/Hz. Fig.~\ref{fig:example2} shows the
impact of channel uncertainty on the worst-case secrecy rate for
different number of Eves. We see in Fig.~\ref{fig:example2} that the
proposed robust design achieves a higher worst-case secrecy rate
than the non-robust design over the whole uncertainty region tested.

This paper has proposed a joint optimization approach to the AN
aided covariances design for the robust secrecy rate maximization
problem. We have shown that the corresponding robust optimization
problem can be handled by performing a one-dimensional line search,
in which a sequence of SDPs are involved. The present work considers a
worst-case achievable secrecy rate problem under deterministic channel uncertainties. As a future work,
it would be interesting to study how this work may be extended to deal with stochastic channel uncertainties, e.g.,
through an outage-based formulation.


\vspace{-10pt}
\section{Appendix}
\vspace{-10pt}
\noindent {\it Proof of Theorem~1:} The proof consists of two steps:
First, we consider a secrecy rate related power minimization (PM)
problem and show that the optimal solution of the PM problem is also
optimal to our main problem~\eqref{eq:rob_SRM}; second, we show that
the optimal ${\bf W}$ of the PM problem has to be of rank one, and
thus establish the existence of a rank-one optimal ${\bf W}$ for
Problem~\eqref{eq:rob_SRM}.

{\it Step 1:} Consider the following power minimization problem:
\begin{subequations} \label{eq:rob_pow}
\begin{align}
\min_{ {\bf W}, {\bm \Sigma} }~~ & {\rm Tr}({\bf W} +
{\bf \Sigma}) \label{eq:rob_pow_a} \\
{\rm s.t.} ~~&  \min_{k=1,\ldots,K} f_{k} ({\bf W}, {\bm \Sigma})
\geq R^\star,~~{\bf W}\succeq {\bf 0}, ~~{\bm \Sigma}\succeq {\bf 0}
\label{eq:rob_pow_b}
\end{align}
\end{subequations}
where $f_k({\bf W},{\bf \Sigma})$ is denoted in \eqref{eq:rob_SRM};
$R^\star$ is the optimal value of Problem~\eqref{eq:rob_SRM}. Here
Problem~\eqref{eq:rob_pow} aims to minimize the total transmit power
given a minimum secrecy rate specification $R^\star$.

Let $(\bar{\bf W}, \bar{\bf \Sigma})$ and $( \hat{\bf W}, \hat{\bf
\Sigma})$ be optimal solutions of Problems~\eqref{eq:rob_SRM} and
\eqref{eq:rob_pow}, respectively. Apparently, $(\bar{\bf W},
\bar{\bf \Sigma})$ is feasible to \eqref{eq:rob_pow}. Thus, we have
that\vspace{-5pt}
\begin{equation}\label{eq:rob_pow_ineq} {\rm
Tr}(\hat{\bf W}+ \hat{\bf \Sigma}) \leq {\rm Tr}(\bar{\bf W}+
\bar{\bf \Sigma}) \leq P,
\end{equation}
which further implies that $(\hat{\bf W}, \hat{\bf \Sigma})$ is
feasible to Problem~\eqref{eq:rob_SRM}, i.e.,
\begin{equation}\label{eq:proof_upper}
\min_{k=1,\ldots,K} f_{k} (\hat{\bf W}, \hat{\bm \Sigma}) \leq
R^\star.
\end{equation}
On the other hand, as an optimal solution of \eqref{eq:rob_pow},
$(\hat{\bf W}, \hat{\bm \Sigma})$ must satisfy \eqref{eq:rob_pow_b}.
Therefore, combining \eqref{eq:rob_pow_b} and
\eqref{eq:proof_upper}, we get
\[  \min_{k=1,\ldots,K} f_{k} (\hat{\bf W},
\hat{\bm \Sigma}) =  R^\star\] i.e., $(\hat{\bf W}, \hat{\bm
\Sigma})$ is also optimal to Problem~\eqref{eq:rob_SRM}.

{\it Step 2:} To prove that the optimal ${\bf W}$ of
\eqref{eq:rob_pow} has to be of rank one, we first re-express
\eqref{eq:rob_pow} as the following problem by using a similar
approach presented in Section~3:
\begin{subequations}\label{eq:rob_pow_SDP}
  \begin{align}
 \min_{{\bf W}, {\bf \Sigma}, \alpha, {\bm \lambda}}~   & {\rm Tr}({\bf W} + {\bf
    \Sigma}) \label{eq:rob_pow_SDP_a}\\
    {\rm s.t.} ~& {\bf h}^H ({\bf W} + (1- \alpha) {\bf \Sigma}) {\bf
    h} +1 - \alpha \geq 0 \label{eq:rob_pow_SDP_b}\\
     ~& {\bf A}_k({\bf W}, {\bf \Sigma}, \alpha, \lambda_k) \succeq {\bf 0},~~k=1,\ldots,K \label{eq:rob_pow_SDP_c}\\
    ~& {\bf W} \succeq {\bf 0},~~{\bf \Sigma} \succeq {\bf 0},
    ~~\lambda_k \geq 0,~ k=1,\ldots,K,\label{eq:rob_pow_SDP_d}
  \end{align}
\end{subequations}
where
\vspace{-5pt} \begin{align} & {\bf A}_k({\bf W}, {\bf \Sigma}, \alpha,
\lambda_k)  = {\bf A}_{k1}(\lambda_k , \alpha) - \bar{\bf G}_k^H
\Big({\bf W} + (1- 2^{-R^\star} \alpha) {\bf
    \Sigma} \Big) \bar{\bf G}_k,\notag\\
    & {\bf A}_{k1}(\lambda_k , \alpha)  = \begin{bmatrix} \lambda_k {\bf I} & {\bf 0} \\
{\bf 0}^H & -\lambda_k \epsilon_k^2 + 2^{-R^\star} \alpha -1
\end{bmatrix},~~\bar{\bf G}_k = \begin{bmatrix}
  {\bf I}, & \bar{\bf g}_k \notag
\end{bmatrix}.
\end{align}
We list part of the KKT conditions of \eqref{eq:rob_pow_SDP} below
\begin{subequations}\label{eq:kkt}
  \begin{align}
    {\bf I} - \eta {\bf h} {\bf h}^H + \textstyle \sum_{k=1}^K \bar{\bf G}_k {\bf
    B}_k \bar{\bf G}_k^H - {\bf Y} & = {\bf 0} \label{eq:kkt_a}\\
    {\bf W} {\bf Y} & = {\bf 0} \label{eq:kkt_b}\\
    {\bf W} \succeq {\bf 0},~ {\bf Y} \succeq {
    \bf 0},~ \eta \geq 0,~ & {\bf B}_k \succeq {0},~\forall k \label{eq:kkt_c}
  \end{align}
\end{subequations}
where ${\bf Y}$, ${\bf B}_k$ and $\eta$ are dual variables
associated with ${\bf W}$, ${\bf A}_k$ and \eqref{eq:rob_pow_SDP_b},
respectively.

Premultiplying \eqref{eq:kkt_a} by ${\bf W}$ and making use of
\eqref{eq:kkt_b}-\eqref{eq:kkt_c}, we have
\begin{equation} \label{eq:kkt_equality1}
{\bf W}\Big ( {\bf I} + \textstyle \sum_{k=1}^K \bar{\bf G}_k {\bf
    B}_k \bar{\bf G}_k^H \Big) =  \eta {\bf W} {\bf h} {\bf h}^H.
\end{equation}
    Therefore, the following relation holds
    \begin{subequations} \label{eq:kkt_equality2}
    \begin{align}
{\rm rank}({\bf W}) & =  {\rm rank} \Big ({\bf W} \big ( {\bf I} +
\textstyle \sum_{k=1}^K \bar{\bf G}_k {\bf
    B}_k \bar{\bf G}_k^H \big) \Big) \label{eq:kkt_equality2_a}\\
    & = {\rm rank}(\eta {\bf W} {\bf h} {\bf h}^H) \leq 1 \label{eq:kkt_equality2_b}
    \end{align}
    \end{subequations}
    where \eqref{eq:kkt_equality2_a} follows from ${\bf I} +
\sum_{k=1}^K \bar{\bf G}_k {\bf
    B}_k \bar{\bf G}_k^H \succ {\bf 0}$; \eqref{eq:kkt_equality2_b}
    follows from \eqref{eq:kkt_equality1}, and the fact that ${\bf h}{\bf h}^H$
    is a rank-one matrix. Since $R^\star > 0$, ${\bf W} = {\bf
    0}$ is infeasible to Problem~\eqref{eq:rob_pow}. Thus, ${\rm rank}({\bf W}) =
    1$ must hold true, which completes the proof.

\vspace*{-5pt}
\bibliographystyle{IEEEtran}
\footnotesize
\bibliography{ref}
\vspace*{-8pt}

\end{document}